**Piotr GAS**
AGH University of Science and Technology

# Essential Facts on the History of Hyperthermia and their Connections with Electromedicine

***Abstract***. *The term hyperthermia is a combination of two Greek words: HYPER (rise) and THERME (heat) and refers to the increasing of body temperature or selected tissues in order to achieve a precise therapeutic effect. This paper reviews the development of thermotherapy by describing the most important moments in its history. For decades, the development of hyperthermia ran parallel with the development of cancer treatment and had numerous connections with electromedicine. Throughout its history, hyperthermia evoked a number of hopes, brought spectacular successes, but also was the subject of many disappointments.*

***Streszczenie***. *Termin hipertermia stanowi połączenia dwóch wyrazów z języka greckiego: HYPER (podnosić) oraz THERME (ciepło) i odnosi się do podnoszenia temperatury ciała lub wybranych tkanek w celu osiągnięcia określonego efektu terapeutycznego. W niniejszej pracy dokonano przeglądu rozwoju terapii ciepłem przez opis najważniejszych momentów w jej historii. Przez dziesięciolecia rozwój hipertermii biegł równolegle z rozwojem leczenia nowotworów i miał liczne związki z elektromedycyną. W swojej historii terapia ciepłem budziła liczne nadzieje, przynosiła spektakularne sukcesy, ale również była przedmiotem wielu rozczarowań.* **(Istotne fakty z historii hipertermii i ich związki z elektromedycyną)**

**Słowa kluczowe**: historia hipertermii, leczenie raka, elektromedycyna

## Introduction

Hyperthermia as a method of treating cancer has a long history, dating back to around 3000 B.C. The term hyperthermia is a combination of two Greek words: *hyper* (rise) and *therme* (heat) and refers to the increasing of body temperature or selected tissues in order to achieve a precise therapeutic effect. Throughout its history, hyperthermia evoked a number of hopes, brought spectacular successes, but also was the subject of many disappointments. Scientists are still looking for new techniques that will make hyperthermia a simpler, safer, more effective and widely available method for patients. The use of nanotechnology in hyperthermia treatment e.g., magnetic fluid hyperthermia, which is currently under experimentation, seems particularly promising [11].

## Heat in the Ancient Times

The use of high temperature as a method of treatment of various diseases (including cancer) was common in various cultures since ancient times. Primarily, the heat had sacral meaning and was associated with the healing power of the Sun [16]. Therefore, it was used for the treatment of locally affected human body parts or the whole organism. For that purpose, hot water and sand (mud baths) from natural thermal springs, and hot air and steam occurring in the volcanic caves were utilized. The first known use of heat treatment was carried out by an Egyptian aruspice named Imhotep (2655 – 2600 B.C.).The Edwin Smith papyrus from about 1700 B.C., which is probably a copy of a thousand year old text, reports that the ancient Egyptians used the so-called "fire drills" (hot blades and sticks) for the treatment of breast cancer [19]. These treatments involved the burning of cancerous cells and had nothing to do with elevated body temperature. It is worth noting that the local and systemic hyperthermia treatments were also very popular in ancient China and India [1].

In ancient Greece and Rome many physicians shared the opinion that knowledge how to control human body temperature will allow them to cure all diseases, including cancer, whose pathology was well known and described in those days [2]. For example, the Greek philosopher Parmenides (ca. 540 – ca. 470 B.C.) was deeply convinced of the effectiveness of hyperthermia as evidenced by the words: "Give me the power to produce fever and I will cure all diseases" [9]. This view was shared by Hippocrates (460 – 370 B.C.), a Greek philosopher and scientist who is considered the "father of medicine". He claimed that the disease must be incurable, if it can not be cured by using heat. Moreover, Hippocrates successfully used heat to treat breast tumors [16]. His medical practice was based on the philosophy of ancient Greece, which attributed the fire to the highest level of intelligence and freedom [17]. Heat treatment was recommended after unsuccessful trials of surgery and when known drugs and other treatments failed. This conviction is well illustrated by the words of Hippocrates: "What medicines to not heal, the lance will; what the lance does not heal, fire will" [20]. Belief in the curative effect of fever was also shared by Celsus (ca. 25 B.C. – 45 A.D.), a Roman author of the first systematic treatise on medicine "De Medicina" and Rufus of Ephesus, a Greek physician who lived at the turn of the 1st and 2nd century [20]. Celsus described the hot baths as a tool in the treatment of various diseases.

## Hyperthermia and infectious fever

The first paper on hyperthermia was published in 1866 by a German surgeon Carl D. W. Busch (1826 – 1881). He described the case of a 43-year-old woman with advanced sarcoma on her face [18]. After the tumor was removed, the patient fell ill with erysipelas. The disease induced high temperature which led to tumor regression. Busch's discovery was fundamental because it was the first reported case showing that high temperature can selectively kill cancerous cells while not affecting the healthy. This event and other similar reports had led to increased interest in hyperthermia and make many trials to induce infectious fever in patients with cancer, for example, by applying dirty bandages or blood of people suffering from malaria into open wounds [20].

Apart from the research of Busch, the relationship between infection and cancer regression was looked for by a young American surgeon William B. Coley (1862 – 1936). His studies involve the injection of different types of bacterial pyrogens into tumors and observation of their behavior. In 1891 he developed a toxin that caused typical erysipelas with its typical fever [3]. For this reason, Coley is called the father of the modern use of hyperthermia and immunotherapy against cancer. So-called Coley'a toxin had been used to treat various types of cancer for nearly a century. Results of research conducted by Coley'a showed that the five year survival rate increased from 28 to 64% for patients with inoperable cancer depending on the temperature of concurrent fever. When the temperature was higher the survival time was longer. However, treatment of an infectious fever had one basic defect – it was unpredictable. Each patient responded to the toxin in a different way! All of this caused that over time this method was abandoned and physicians looked for other more effective cancer treatments, which gave more reproducible results.

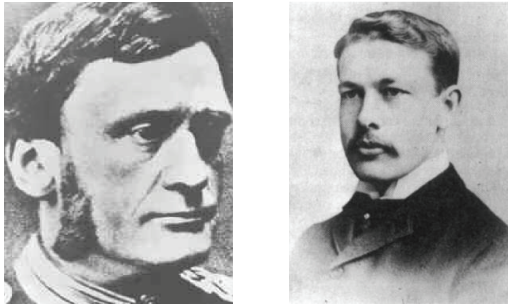

Fig.1. Carl D. W. Busch (left) and William B. Coley (right)

**Local and systemic hyperthermia**

Irrepressible attempts of creating a high body temperature led to the development of various methods of systemic hyperthermia. They mainly involved wrapping the patient in plastic and dipping in hot wax or placing in a specially heated rooms or boxes. More sophisticated techniques required the removal of the patient's blood and forcing it into the body then after it had been warmed (perfusion hyperthermia). These all treatments were characterized by variable success and often led to death. Nevertheless, utilization of hyperthermia had been widely reported in the treatment of sexually transmitted diseases, neurological disorders, arthritis and asthma [20].

Simultaneously, the local hyperthermia was used. In this type of treatment specific tissues or tumor were heated. For example, a Swedish doctor Frans J. E. Westermark (1853 – 1941) in 1898 attempted to treat inoperable cancers of the uterine cervix by using high temperature water (42 – 44$^o$C) circulating in a special metal coil [4]. Effectiveness of the treatment (ongoing for 42 hours) was confirmed by the fact that among the seven patients treated by Westermark, in one case the tumor had disappeared, and in other cases improvement was observed. However, due to extreme pain and long treatment this method was abandoned soon.

**Heating and early electricity**

The first attempts of cancer treatment using electricity were possible after 1800, when the Italian physicist and physiologist Alessandro G. A. A. Volta (1745 – 1827), inspired by the experiences of his compatriot Luigi Galvani (1737 – 1798), invented the electric cell, later called voltaic pile. This primary form of the electric battery was a source of direct current. A single electric pile could generate only a small voltage of 1 – 2 V, but several piles connected in series could create a high power energy source [13]. Voltaic pile quickly became known worldwide and inspired many scientists to do further research and relevant experiment relevant for the development of electricity and medicine.

At that time there was considerable competition among researchers for the recognition of the priority of use of electricity for heating tissue. Early treatment involved burning the tumors by a direct passage of electrical current and was called galvanocautery. Electricity was applied directly to the diseased tissues using special needles, wires and knife-type electrodes (so-called "electric knife") [8]. Already around 1830, a French surgeon and gynecologist Joseph C. A. Recamier (1774 – 1852) used electric heat to treat of uterine cancers [16]. A much more important role in the use of heat in the women's oncology was played by American surgeon John Byrne (1825 – 1902). He was known for the invention of a special liquid storage battery, which supplied sufficient electrical current to cauterize pathological changes of the uterus [14]. Results of his 20-year study were compiled in a text from 1889, in which he summarized the effects of treatment of the uterus and cervix tumors in 367 patients [16]. He confirmed the known evidence that "deeper lying cancer cells are destroyed by less heat than will destroy normal tissues". Similar treatments in men were carried out by the Italian surgeon an Enrico Bottini (1835 – 1903), who in 1874 invented a special device for firing prostate cancer cauterization named as "cauterio termogalvanico" [15].

**Heating and high-frequency currents**

After the English physicist James C. Maxwell (1831 – 1879) formulated his theory of electromagnetism in 1873, and a German physicist Heinrich Hertz (1857 – 1894) confirmed it in 1886 by constructing an oscillator of electromagnetic waves (so-called the Hertz oscillator), the rapid development of many types of AC devices was imminent. One person, who drew attention to the possibility of using high-frequency currents for therapeutic applications, was a physicist Nikola Tesla (1856 – 1943). His publications in the "Electrical Engineer", referred not only to the physiological effects, but also to the heating capability of tissues using alternating current above a certain frequency. Interestingly, Tesla was miraculous cured of pulmonary tuberculosis by the high-frequency currents in 1899 [8].

However, the most famous person who established the ground for capacitive and inductive heating of tissues using high-frequency currents was a French physicist and physiologist Jacques-Arsene d'Arsonval (1851 – 1940). Already in 1981 he carried out a significant experiment in which he passed a current of 3A through his own body. What was important, he did not feel any sensation with the exception of heat [16]. He experienced more heating effects during a similar experiment in 1898, when current of 500mA was passed through his body. As a result of these experiments he introduced what he called "autoconduction" which was really induction under the influence of high-frequency electromagnetic field [7]. Since that time, he commenced to employ high-frequency currents in the treatment of various diseases.

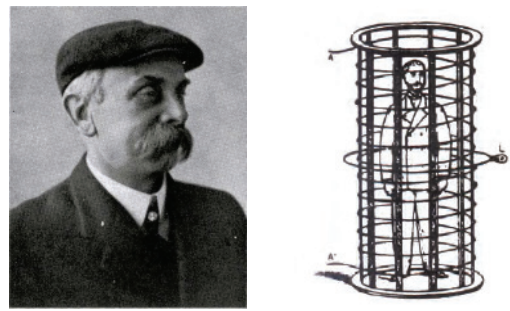

Fig.2. Prof. d'Arsonval and his induction coil applicator

During the years 1894 – 1895 d'Arsonval clinically treated 75 patients suffering from various ailments. Each of them was placed in the huge induction solenoid (Fig.2) and was exposed to the fields for 15 – 20 minutes a day. Other patients were treated on a so-called condenser couch (Fig. 3), where currents of 350 – 450 mA were passed through their body for 6 minutes a day [8]. These devices created by d'Arsonval, enjoyed great popularity at the turn of the century [10] and the so-called "Arsonvalisation" have been widely used in the treatment of many diseases, beginning from disorders of blood circulation, metabolism, central nervous system and skin, and ending with tuberculosis, gastrointestinal disorders and hemoroids. It was also claimed that Arsonvalisation had positive influence on the central thermoregulatory system of the body [16].

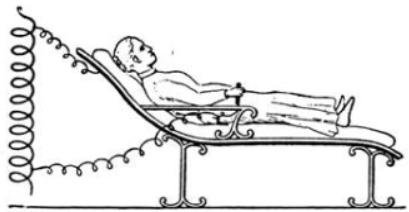

Fig.3. D'Arsonval's condenser couch for capacitive heating

The actual mechanisms of heating using high-frequency current were explained by von Zeyneck in 1899. He believed that the heat is generated by the passage of current through the tissue similarly as when the resistance is passed by electric current [8].

In 1907 a German physician Karl Franz Nagelschmidt (1875 – 1952) demonstrated the possibility of deep heating of tissue using high frequency energy, and suggested its use in the treatment of blood circulation and joints diseases. In 1909 he introduced the concept of diathermy to determine the heating effects inside tissues (so-called "thermooperation") caused by high frequency currents [16]. His book on diathermy of 1913 initiated a new era of high-frequency electrotherapy applications (Fig.4). Independently from Nagelschmidt, the heat tissue penetration (so-called "thermopenetration") was described in 1909 by an Austrian physician Gottwald Schwarz (1880 – 1959), and in 1910 a German physician Christoph Müller performed it with diathermy. This treatment involved the electrical heating of the tissue without its destruction in order to increase blood flow [2].

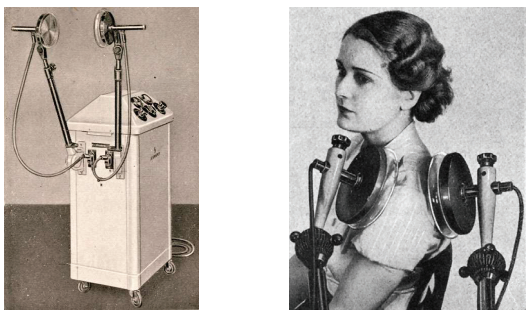

Fig.4. Apparatus for diathermy and possible way of its application

The concept of diathermy contributed to the mistaken belief that the biological effects of electromagnetic fields were limited only to thermal effects. The development of powerful generators in the following years led to the creation of new types of diathermy techniques their applications in medicine. With time, apart from previously known longwave diathermy (0.5 – 3.0 MHz), shortwave diathermy with frequencies of up to 100 MHz by 1920 and shortwave diathermy with frequencies of 100 – 3000 MHz by 1930 became standard tools in medical treatments. This type of heating technology and high-frequency currents were eventually introduced into the surgery practice (surgical diathermy) [16].

In 1927, Swedish gynecologist Nils Westermark (1892 – 1980), the son of F. Westermark, published in his dissertation the results of experimental studies on the heating of tumors in rats with high frequency [4]. He said that the tumors may be more sensitive to heat than healthy tissues and also recommended the temperature of 40 – 45°C for selective tumor heating. Shortwave diathermy was first used in 1928 by a German physician Erwin Schliephake (1894 – 1995) to cure a furuncle on his nose [8]. He initially investigated the biological effects of shortwave treatment on various tissue types and then introduced it to clinical applications.

**Hyperthermia and other therapies**
The discovery of ionizing radiation in 1895 by German physicist Wilhelm C. Röntgen (1845 – 1923) led to the use of X-rays to treat cancer and temporarily reduce the interest in hyperthermia in the early twentieth century. Due to insufficient and underdeveloped heating methods and temperature measurement techniques, as well a the lack of positive clinical results, the consent to hyperthermia treatment with hyperthermia had been withdrawn. Since then cancer was dominated by traditional surgical methods as well as radiotherapy and chemotherapy [18].

However, already in 1910, G. Schwarz found that hyperthermia combined with radiotherapy gives good therapeutic results. In 1912, Müeller discussed in detail the clinical effects of radiotherapy combination with hyperthermia and presented the treatment results of 100 patients with various tumors treated with long-term therapy [2]. In 1/3 of cases the tumors had disappeared, and another third part showed an improvement though with subsequent tumor regrowth or metastases. Ch. Müller recommended association treatment, but warned that "the deeper the tumor location, the fewer options there are for treatment success". Unfortunately, from the viewpoint of modern science that study was not a random clinical trial.

After 1929 E. Schliephake developed a set of paired electrodes to induce capacitive heating in advanced cases of uterine cancer. He used radio frequencies of 20 – 50 MHz with wavelengths of 6 – 15 m for different heating periods of up to 40 minutes. He noticed improvement in tumor tissue destruction when heat treatment was associated with X-rays [16]. With time this method was modified and used for the treatment of various pelvic tumors. Although Schliephake's method was successful in the treatment of inoperable cases, it did not find its way into mainstream of medicine.

The revival of interest in hyperthermia occurred after World War II, when it research on the biological reasons for hyperthermia began. In 1962 an American surgeon, George W. Crile Jr.. (1907 – 1992) made a famous discovery, stating that long-lasting increase of the temperature of some tumors to 42 – 50°C could selectively destroyed them without damaging the healthy tissues [18]. In 1971 American researchers A. Westra and W. C. Dewey made the first serious experiment on hyperthermia of mammalian cells. It became an inspiration to the study radio- and chemo-sensitivity of tumor cells after hyperthermia treatments. Soon, more effective heat treatment in combination with radiotherapy was confirmed E. Ben-Hur. In 1977, Dewey proved that long-term (over 30 minutes) exposure of cells to temperatures exceeding 40°C resulted in their death [5]. Moreover, it was noted that at temperatures above 42.5°C, a slight increase in temperature (of the order of tenths of a degree) caused a significant increase in cell mortality. These results signalled the importance of accurately determining the temperature in hyperthermia systems, and precision of the temperature measurement within the tumor. At the same time techniques for perfusion hyperthermia with anticancer drugs for the treatment of tumors in isolated organs (liver, lungs) and limbs had been developed and standardized [12].

**Microwave Hyperthermia**

A technological revolution in hyperthermia occurred when the microwave heating techniques were developed and were applied in medicine. The development of new technologies allowed for better options in the local heating of tumors, including those located deep in the human body (local-regional hyperthermia). During the period of 1938 – 1939 H. E. Hollman discussed the possibility of therapeutic applications of 25cm waves; he predicted that the waves could be focused to produce the heating of the deep tissues without excessive heating of the skin. In 1946 the Federal Communications Commission assigned the frequency of 2450 MHz to physical medicine based on its alleged superiority in therapeutic value [8]. Next L. L. Siems performed comparative studies of the effect of shortwave and microwave diathermy on blood flow in 1948. Two years later A. E. Gessler and partners probably were the first group of physicians to use RF energy at microwave frequencies in the experimental treatment of cancer. In the years 1977 – 1983 a Swedish oncologist Jens Overgaard, carried out 204 separate clinical studies with 10,265 participating patients, which confirmed the effectiveness of thermotherapy in cancer treatment [4]. Since that time hyperthermia has been considered a standard method cancer therapy .In 1990 Alan J. Fenn, an electrical engineer at the MIT's Lincoln Laboratory, developed a concept for heating deep tumors by means of adaptive microwaves. These adjust to the properties of a patient's tissue to concentrate the microwave energy at the tumor position. These adaptive microwaves are generated by multiple microwave antennas (an adaptive phased array) that surround the human body [6].

**Modern Hyperthermia**

Worldwide interest in hyperthermia had become the starting point for organizing the International Symposium on Cancer Therapy by Hyperthermia and Radiation in Washington, D.C. (USA) in 1975. Other similar congresses took place in Essen (Germany) in 1977, and in Fort Collins (Colorado, USA) in 1980. Since then, the International Symposium on Hyperthermic Oncology has been held every four years. In 1981 the North American Hyperthermia Society was created and two years later the European Hyperthermia Institute was formed. In Japan, hyperthermia research started in 1978 and the Japanese Society of Hyperthermia Oncology was established in 1984. In 1985 the North American Hyperthermia Society, together with the European Society for Hyperthermic Oncology and the Japanese Society of Hyperthermic Oncology cooperatively founded the International Journal of Hyperthermia and adopted it as their official journal. Currently, it is the most important periodical devoted to hyperthermia. In 1988, by combining hyperthermia with modern knowledge of bio-electromagnetism and human physiology, a new direction in the fight against cancer called "oncothermia" was created, which is still being developed [17]. It is worth noting that in 2010 the First International Symposium Oncothermia took place in Cologne (Germany) and the first issue of the Oncothermia Journal was published.

**Summary**

Nowadays hyperthermia is an emerging effective treatment method in oncology. Against its long history hyperthermia is still in its infancy. Positive therapeutic effects of this treatment depend on the applied temperature, exposure time and the volume of the tissue exposed to electromagnetic fields. The effectiveness of heat treatment can be significantly increased by combining hyperthermia with other cancer treatments such as radiotherapy, chemotherapy, immunotherapy, and gene therapy. Hyperthermia has found application in the treatment of breast, skin, head and neck, brain, oesophagus, prostate, uterus cervix, and urinary bladder tumors. However, still not all mechanisms responsible for cell death during hyperthermia are known. New methods of heating are being studied all the time. Many opportunities presented itself with the development of nanotechnology and magnetic fluid hyperthermia observed in recent years.

*Authors*: mgr inż. Piotr Gas, AGH University of Science and Technology, Department of Electrical and Power Control Engineering, al. Mickiewicza 30, 30-059 Kraków, E-mail: piotr.gas@agh.edu.pl